\documentclass[fleqn,10pt]{wlscirep}
\usepackage[utf8]{inputenc}
\usepackage[T1]{fontenc}
\usepackage{mathtools}
\title{Impact of spin-orbit coupling and Zeeman interaction on the multiple Andreev reflections subharmonic gap structure in nanoscopic Josephson junctions}

\author[1,*]{Dibyendu Kuiri}
\author[1]{Jorge Huamani Correa}
\author[1]{Andrzej Biborski}
\author[1, +]{Michał Piotr Nowak}
\affil[1]{AGH University of Krakow, Academic Centre for Materials and Nanotechnology, al. A. Mickiewicza 30, 30-059 Krakow, Poland.}

\affil[*]{kuiridibyendu9547@gmail.com}
\affil[+]{mpnowak@agh.edu.pl}

\begin{abstract}
Multiple Andreev reflections in voltage-biased Josephson junctions give rise to the subharmonic gap structure in the conductance, which is widely used to characterize transport properties and estimate the induced gap in the junctions. Here we theoretically investigate the evolution of the subharmonic gap structure in spinful Josephson junctions. Spin mixing introduced by the spin-orbit coupling opens avoided crossings in the dispersion relation of the leads, which, as we demonstrate, subsequently results in pronounced multiple Andreev reflection features in the conductance traces. We analyze how these features evolve under an external magnetic field and explain that their visibility in conductance is governed by the spin polarization of the bands.
\end{abstract}
\begin{document}

\flushbottom
\maketitle
%
%
\thispagestyle{empty}


\section*{Introduction}
Josephson junctions under the application of external voltage bias exhibit specific current- and conductance-voltage responses with pronounced features at voltages corresponding to the integer fractions of the superconducting gap, $V_b = 2\Delta/en$. This phenomenon was first observed in superconducting point contacts \cite{PhysRev.184.434, PhysRevLett.31.524} and tunnel junctions \cite{PhysRev.172.393}. Soon after those findings, the term multiple Andreev reflections (MAR) has been coined \cite{KLAPWIJK19821657} which describes a process that stands behind the observed subgap structure: for voltages below twice the superconducting gap, quasiparticles incident on a superconducting lead undergo sequential Andreev reflections, the number of which depends on the applied voltage, thereby transferring multiple Cooper pairs across the junction [see Fig. \ref{fig:system}(c)].

Further theoretical development allowed the description of junctions with arbitrary transparencies \cite{PhysRevB.27.6739, PhysRevLett.74.2110} and the coherent regime, including both AC and DC components of the current, in single-channel \cite{PhysRevLett.75.1831}, multimode \cite{PhysRevB.56.R8518} and multiterminal \cite{PhysRevB.95.075417, PhysRevB.99.075416, PhysRevB.65.134523} cases. The predictions of the theory of Ref. \cite{PhysRevLett.75.1831} were confirmed with impressive accuracy in break junctions \cite{PhysRevLett.73.2611, PhysRevLett.78.3535, PhysRevB.56.R8518}, where the analysis of MAR conductance traces allowed extraction of the number of modes and the transparency of the junction. The ability to determine both the number of quantized modes and their transmission probabilities, as well as the size of the induced gap, has become a powerful tool for characterizing novel semiconductor-superconductor hybrids \cite{Goffman_2017, deVries2018, PhysRevApplied.7.034029, Borsoi2021, Heedt2021, adma.202403176, Salimian2021} where MAR are observed \cite{Gunel2012, Nilsson2012,PhysRevMaterials.3.084803}.

Hybrid semiconductor–superconductor devices, particularly those with large g-factors and strong spin–orbit interaction (SOI), have been intensively studied for engineering topological superconductivity and realizing Majorana bound states \cite{PhysRevLett.105.077001,Oreg2010,Mourik2012,Das2012,Deng2016,Bommer2019}. MAR have been considered as a tool to trace the topological transition \cite{San-Jose_2013} or chirality of Andreev bound states \cite{PhysRevB.107.184510} in topological junctions. The interplay between the Zeeman interaction, SOI and superconductivity \cite{Pang2025} leads to a complex energy structure of the proximitized semiconductor. Already, experiments on quasi-one-dimensional nanowires revealed complicated MAR conductance spectra when the magnetic field was applied \cite{Heedt2021}. Furthermore, a recent experiment demonstrated that, in a ferromagnetic proximitized nanowire, additional subgap peaks appear in the conductance spectrum of the system \cite{k8sv-lp1j} due to the Zeeman shifting of the gaps opened in the dispersion relation of the leads, which can modify the MAR spectrum \cite{PhysRevB.101.020502}. 

The aim of this paper is to establish the interplay between the energy structure of the superconducting leads with the conductance response of a biased Josephson junction. We relate the gaps in the dispersion relations to features in the conductance spectra and connect them to spin selection rules that govern quasiparticle transport. The established methods to theoretically capture MAR range from simplified wave-function matching techniques based on the scattering matrix of the normal region \cite{PhysRevLett.75.1831} to more complex Hamiltonian approaches \cite{PhysRevB.54.7366}. Here we rely on a direct coherent microscopic description of the time-dependent current transport through the junction. We consider quasi-one-dimensional junctions, explicitly taking into account the Rashba spin-orbit coupling, Zeeman interaction, and opaque scattering region, and via time evolution we obtain the dissipative quasiparticle current. We directly relate the evolution of the superconducting lead's band structure to features in the junction conductance, emphasizing the impact of the gaps that arise from the interplay of SOI and Zeeman interaction on the subharmonic MAR structure.

\section*{Model}
\begin{figure}[h]
    \centering
    \includegraphics[width = 0.8\textwidth]{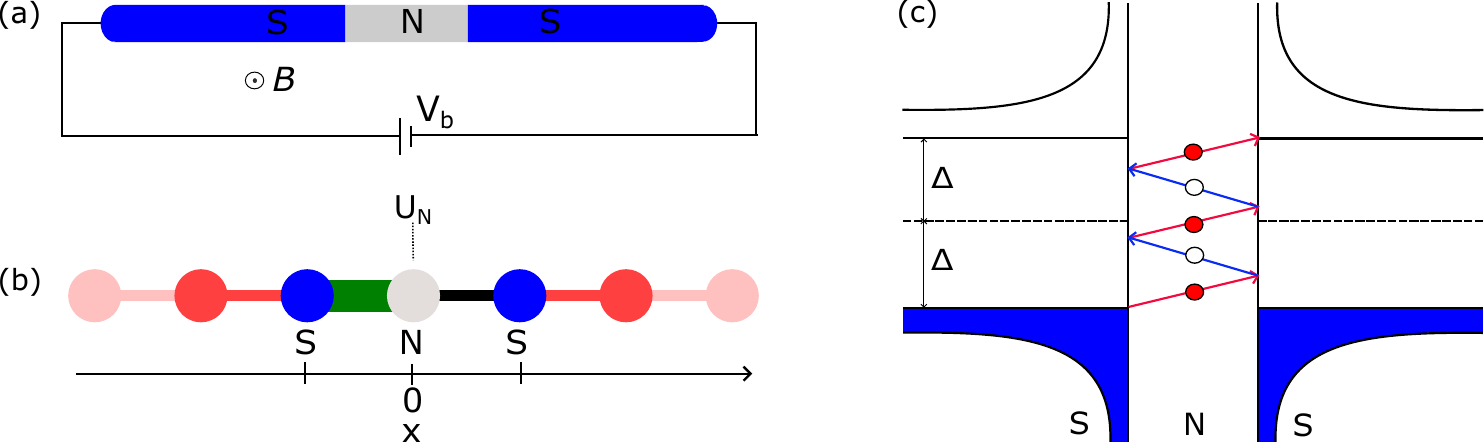}
    \caption{(a) Schematic of the system: a nanowire with superconducting regions (S) on both sides, separated by a normal region (N) biased by the voltage $V_b$ and under the application of perpendicular magnetic field. (b) Numerical setup: red circles indicate semi-infinite superconducting leads, while gray and blue circles correspond to the normal and superconducting regions, respectively. Black and red lines denote standard hopping, while the green thick line represents hopping with the phase shift induced by the time-dependent superconducting phase difference $\phi(t)$ due to voltage bias. (c) A schematic illustration of scattering processes in a voltage-biased Josephson junction. For bias voltages smaller than $2\Delta/e$, the incident particle undergoes sequential Andreev reflections (with red circle denoting electron and blue the hole) before its energy becomes sufficient to surpass the superconducting gap.}
    \label{fig:system}
\end{figure}

Here, we consider Josephson junctions based on a single-channel semiconducting nanowire with proximity-induced superconductivity described by the Hamiltonian: 
\begin{equation}\label{eq:1}
\begin{split}
    H = &\left( \frac{\hbar^2 \hat{k_x}^2}{2m^*} -\mu  + U_N(x)\right)\sigma_0 \otimes \tau_z 
    + E_z \sigma_z \otimes \tau_0  - \alpha \sigma_y \hat{k_x} \otimes\tau_z + \Delta(x,t) \sigma_0\otimes\tau_{+} + \Delta^{*}(x,t) \sigma_0\otimes\tau_{-} , 
\end{split}
\end{equation}
written in the basis of $\Psi = (\psi_{e\uparrow}, \psi_{h\downarrow}, \psi_{e\downarrow}, -\psi_{h\uparrow})^T$ (where $e$ and $h$ correspond to electron and hole components with spin up $\uparrow$ or down, $\downarrow$, respectively). $\hat{k_{x}} = -\iota \partial/\partial x$, $\sigma_i$, and $\tau_i$ with ($i = x, y, z$) are the Pauli matrices that act on spin and electron-hole degrees of freedom, respectively, where $\sigma_0$ is ($2 \times 2$) identity matrix. 
The operators $\tau_{+}$ and $\tau_{-}$ are defined as $\tau_{\pm} = \tfrac{1}{2}(\tau_x \pm \iota \tau_y)$.
$\alpha$ is the strength of the SOI present in the nanowire, and we consider a magnetic field applied perpendicular to the nanowire with the Zeeman interaction strength given by $E_z = g\mu_b B/2$.

We discretize the Hamiltonian on a discrete mesh with lattice constants $a = 5$ nm. The system consists of two semi-infinite proximitized segments and a scattering region containing three sites---see Fig. \ref{fig:system}(b). The central part is the normal region where a potential barrier $U_N(x)$ is introduced to control the transparency of the junction. For each current calculation, we determine the necessary value of the potential $U_N$ that ensures the desired transparency $D$ of the normal region by obtaining the transmission probability through the scattering region in the Kwant package \cite{Groth2014_Kwant}. We particularly focus on the case of a junction with limited transparency $D = 0.2$ where the position of the peaks in the conductance trace is related to the relative distance between the superconducting gap edges \cite{PhysRevApplied.7.034029}.

The superconducting pairing potential $\Delta(x,t)$ is defined as 
\[ \Delta(x, t) = \begin{cases*}
                    \Delta_{0} & if  $x > 0$  \\
                     \phantom{}0 & if $x =0$\\
                     \phantom{}\Delta_{0}e^{-\iota\phi(t)} & if $x <0$.
                 \end{cases*} \]%
The step-function form of $\Delta(x,t)$ is justified by the fact that the intrinsic pairing interaction exists only in the parent superconductor, with the semiconductor acquiring superconductivity through the proximity effect. Here, for a voltage-biased Josephson junction, the superconducting phase difference is inherently time-dependent and defined as $\phi(t) = (2e/\hbar)\int_{0}^{t}V_b(t')dt'$ and is taken as:
\begin{equation}
\phi(t) = \begin{cases*}
    V_b\left(t - \dfrac{\tau}{\pi} \sin\left(\dfrac{t\pi}{\tau}\right)\right) & \text{if $t < \tau$} \\
    2V_b\left(t - \dfrac{\tau}{2}\right) & \text{otherwise}.
\end{cases*}
\label{eq:voltageramp}
\end{equation}
The above form of $\phi(t)$ accounts for a voltage ramp during the initial time $\tau$, and then a constant voltage, which results in the linear phase evolution in time.

We follow the approach Ref.~\cite{PhysRevB.93.134506} and the time-dependent pairing term is gauged into a phase-hopping term in the region of $x < 0$ [see Fig.~\ref{fig:system}(b)], i.e., in the hopping between the left superconductor site and normal region site, $t_{i,i-1} \rightarrow t_{i,i-1}\cdot\exp(-\iota\phi/2\tau_z)$. The current is calculated as $I_{i-1 \to i}(t) = 2\,\mathrm{Im}\sum_\gamma\int \tfrac{dE}{2\pi}\, f(E)\,\psi_{\gamma,E}^*(t,i)\,\sigma_0\otimes\tau_z\mathcal{H}_{i,i-1}(t)\,\psi_{\gamma, E}(t,i-1)$, where $f(E)$ is the Fermi function, $\gamma$ enumerates conducting channels, and $\mathcal{H}_{i,i-1}(t)$ is the Hamiltonian matrix element between the sites $i-1$ and $i$. The time-dependent numerical calculations are performed using the algorithms described in Ref. \cite{Kloss_2021} and implemented in the Tkwant package. Despite its simplicity, the adopted model allows for a microscopic description of the junction and MAR processes without resorting to approximations such as the presence of a short junction or the Andreev limit.

For our numerical simulation, we set $\tau = 1000$ $\hbar/eV$, which introduces a ramp-up of voltage at the beginning of the simulation, stabilizing the time evolution. We adopt material parameters typical for InSb proximitized by high critical magnetic field superconductors as Nb, i.e., the induced superconducting gap $\Delta_0 = 2\,\text{meV}$, $m^* = 0.014m_e$, $\mu = 10$~meV, $\alpha = 50$~meVnm, $g = 50$. The code used for the calculation is available in an online repository \cite{kuiri_2026_19204942}. All calculations in this work are performed at zero temperature $T = 0$---as we checked, the elevated temperature results in softening the subharmonic peaks in the conductance but does not affect the processes discussed in this work.

\section*{Results}

\subsection*{Without spin-orbit coupling}
\begin{figure}[ht!]
    \centering
    \includegraphics[width = 0.45\textwidth]{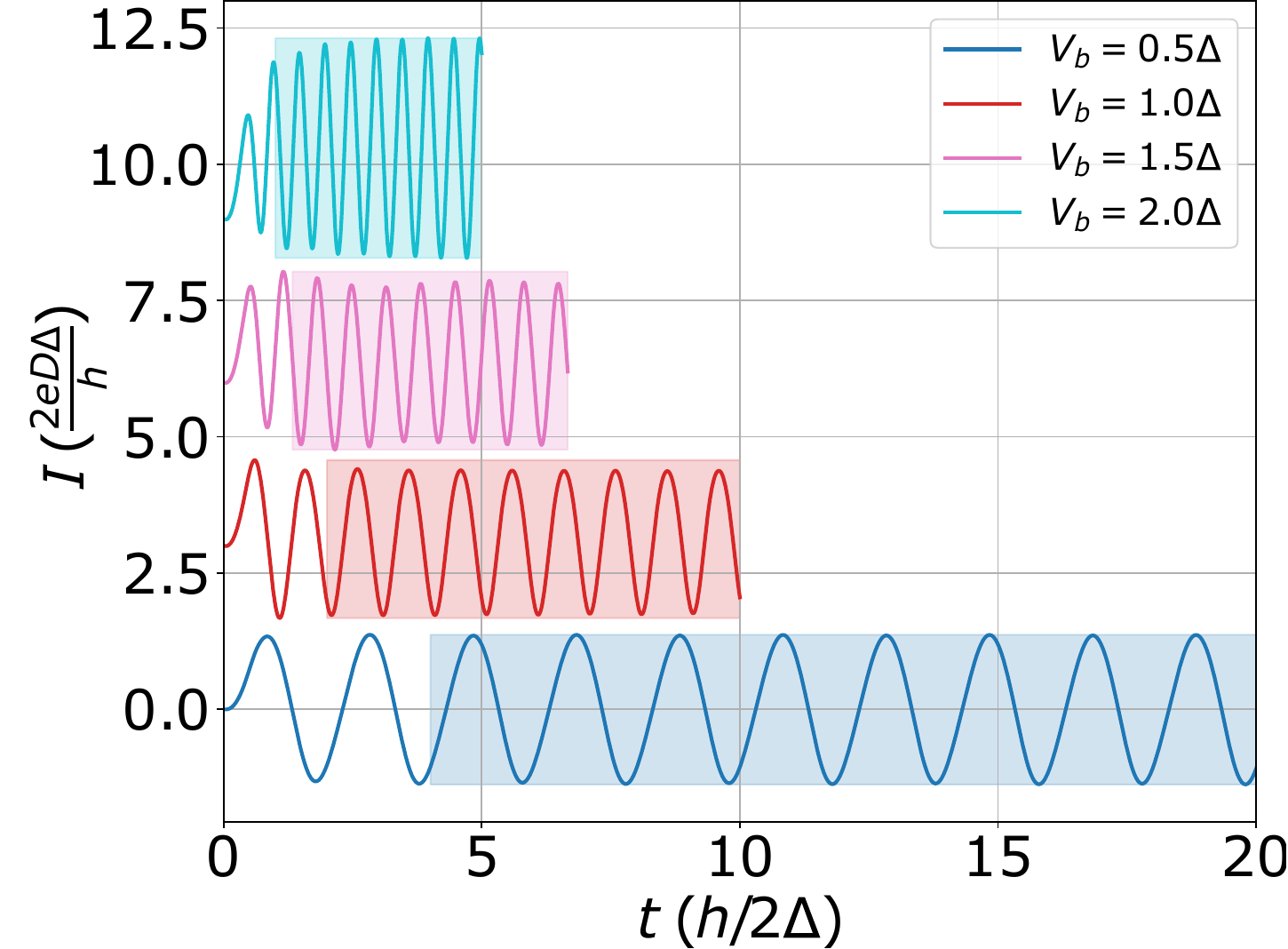}
    \caption{Time evolution of the current $I(t)$ at $B = 0.01~\text{T}$ and $\alpha = 0$, for different voltage biases $V_b = 0.5\Delta, 1.0\Delta, 1.5\Delta,$ and $2.0\Delta$. The shaded regions indicate a typical averaging window $8 t_0$ over which the current is calculated. Curves are plotted with an additional fixed vertical offset of $3\cdot2eD\Delta/h$ along the y-axis.}
    \label{fig:I_t}
\end{figure}

\begin{figure*}[htp!]
    \centering
    \includegraphics[width = 1.0\textwidth]{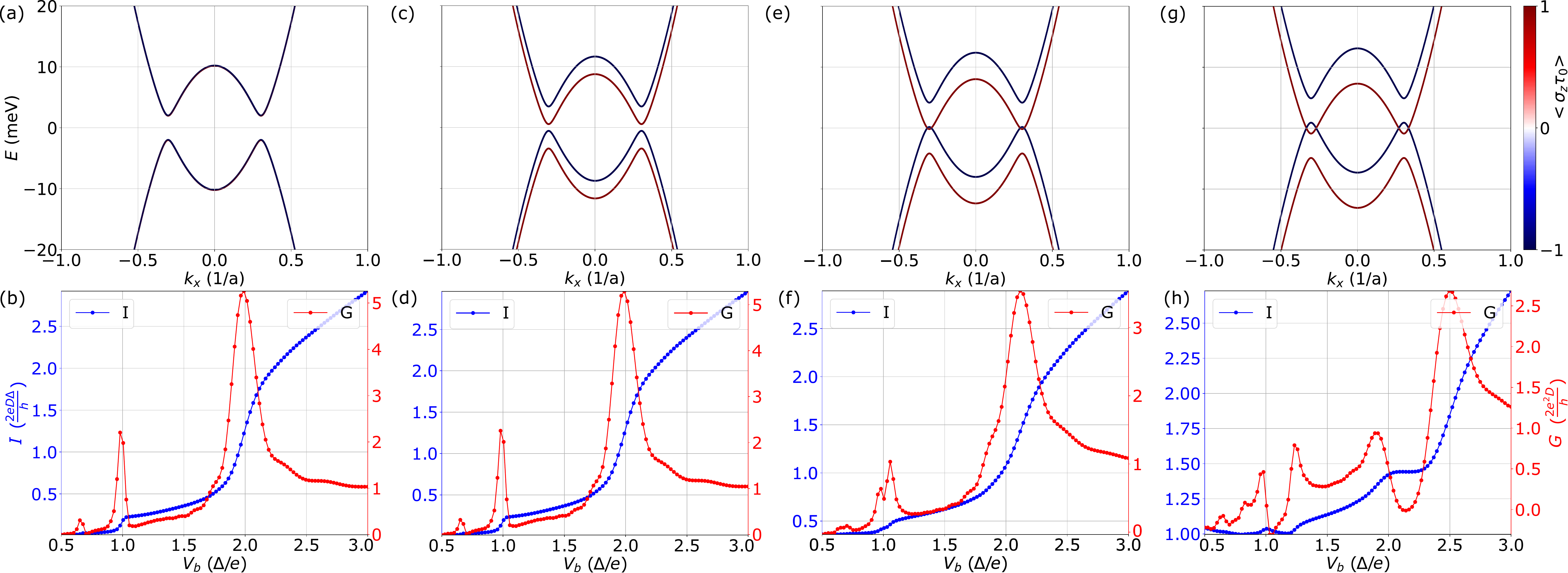}
    \caption{(a), (c), (e), and (g) show the dispersion relation colored by spin polarization for increasing magnetic field and (b), (d), (f), and (h) represent the $I$ vs $V_b$ and $G$ vs $V_b$ for  $B=$ 0.01, 1.0, 1.5, and 2.0 T, respectively.}
    \label{fig:without_soc}
\end{figure*}

Let us start with the case without spin-orbit coupling. The time evolution is carried over a time span determined by the AC Josephson period $t_0 = h/2eV_b$. The resulting time-dependent current obtained for a few values of voltage bias is shown in Fig. \ref{fig:I_t} for a negligible value of the magnetic field $B = 0.01$ T \cite{B0comment}. We allow the time evolution to last for $10t_0$ and use the time span of the last $m$ periods (marked with color overlays for $m=8$) to obtain the time-averaged current after the oscillations stabilize past the initial voltage ramp due to Eq.~\ref{eq:voltageramp}.

Figure \ref{fig:without_soc} in panels (b), (d), (f), (h) shows in blue the averaged current for the increasing magnetic field $B = 0.01$~T, $B = 1$~T, $B = 1.5$~T, and $B = 2$~T respectively. We observe that for small magnetic fields $B= 0.01$ T and $1$ T the current increases in a step-like fashion as the voltage bias is increased. This is a hallmark of the MAR process in which the increase of the voltage reduces the number of Andreev reflections and resulting consecutive transport processes of quasiparticles through the opaque normal region \cite{PhysRevApplied.7.034029}. As a consequence, the conductance trace calculated as $G = dI/dV_b$ and shown in the bottom panels of Fig.~\ref{fig:without_soc} exhibits pronounced peaks. They appear at integer fractions of the superconducting gap $V_b = 2\Delta/en$, characteristic for junctions of reduced transparency. In the considered voltage range, we observe three main peaks at $V_b = 2\Delta, \Delta, 2\Delta/3$. They appear at the voltage values that correspond to quasiparticle energy gain equal to integer fractions of the superconducting gap, which allows for quasiparticles escaping one superconducting lead from the edge of a filled band to reach the other superconducting lead exactly at the edge of an empty band. Interestingly, as the magnetic field is increased from $B= 0.01$~T to $1$~T, both the current and conductance traces remain unchanged---compare Fig.~\ref{fig:without_soc}(b) and (d). 

Let us consider the band structure in the superconducting leads. In the absence of SOI, the system Hamiltonian has a block diagonal form and consists of two separate sets of bands: one for electron and hole pair ($\psi_{e\uparrow},\psi_{h\downarrow}$) and another for ($\psi_{e\downarrow}, -\psi_{h\uparrow}$) pair. The magnetic field, through the Zeeman effect, causes each set to move in the opposite direction in energy as the magnetic field is increased. This becomes clear when comparing Figs. \ref{fig:without_soc}(a) and (c) where the colors denote the two pairs of electron and hole bands with mutually opposite spin orientations. As without the SOI, the two sets of bands are fully decoupled the effective distance between the gap edges in each spin sector remains constant, and thus the MAR conductance and current trace remain unaffected by the Zeeman effect. 

However, this holds only until in the increasing field the bands defining the superconducting gap start to cross the Fermi energy (zero energy here). Figure. \ref{fig:without_soc}(e) shows the band structure for $B = 1.5$ T. Here already some small parts of the bands appear below (above) zero energy for bands colored in red (blue) effectively making this section of the band structure filled (empty) with quasiparticles. This, in a noticeable way, deviates the MAR subgap structure as seen in Fig. \ref{fig:without_soc}(f). The peaks are shifted to larger biases as now it is not the superconducting gap that determines the energy necessary for the decrease of the number of Andreev reflections, but rather the distance between the Fermi energy and the gap edge above---which increases as the magnetic field becomes larger for the bands with a negative value of $\langle \sigma_z\tau_0 \rangle$ (blue in Fig. \ref{fig:without_soc}(e)). 

When we further increase the magnetic field, the MAR subgap features are shifted to even larger bias values, as seen in Fig. \ref{fig:without_soc}(h). Notably, we observe that the conductance increases and has a sudden drop around $V_b = 2\Delta/e$. This can be explained as follows: for the quasiparticles originating from the bands colored red in Fig. \ref{fig:without_soc}(g), the transport from the bottom band edge is allowed for voltage bias $V_b < 2\Delta/e$. Quasiparticles from the top edge of the bottom band gain energy $eV_b$ and, despite having energy below the Fermi level, are Andreev reflected from the gap existing at the energy $E<0$. However, when the bias exceeds $2\Delta/e$, the energy of the particle arriving at the right contact is greater than the gap but also smaller than $0$, so the quasiparticle reaches a filled band and thus cannot be absorbed in that lead. This results in a sudden stop in the current increase as the voltage crosses $V_b = \simeq 2\Delta/e$, which is accompanied by a drop in the conductance. The same process applies to a scattering that involves two Andreev reflections, resulting in a peak at $V_b = \Delta/e$. In addition, in the conductance trace, there is a peak at $V_b \simeq 2.5\Delta/e$ (with its subharmonic partner) that corresponds to the increased distance between the Fermi energy and the gap edge of the band with $\langle \sigma_z\tau_0 \rangle < 0$ colored blue in Fig. \ref{fig:without_soc}(g).

\subsection*{Impact of Rashba SOI}
\begin{figure*}[ht!]
    \centering
    \includegraphics[width = 1.0\textwidth]{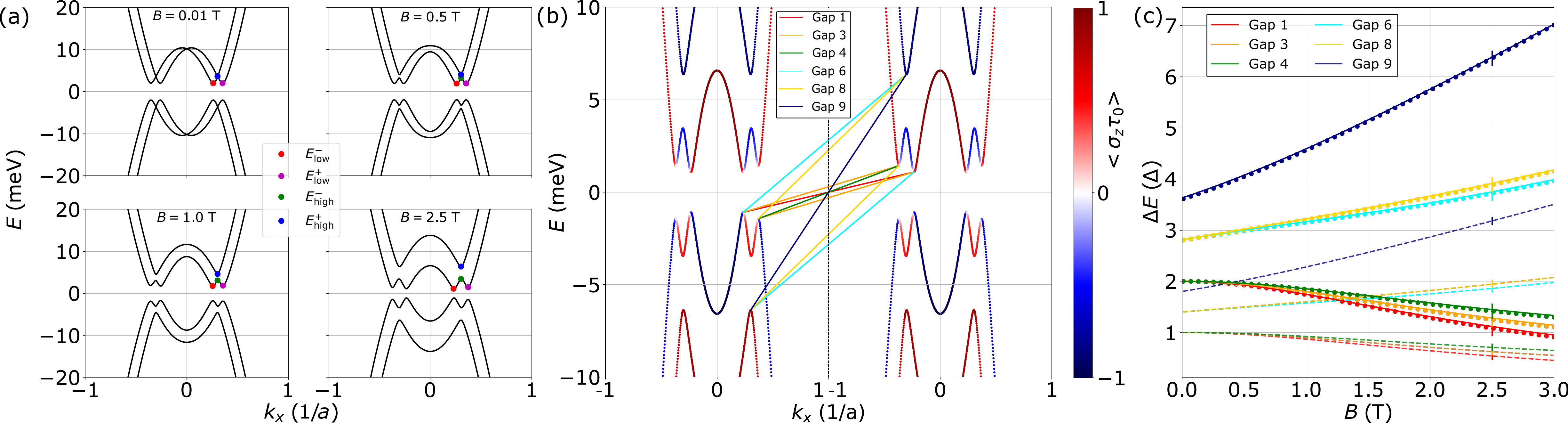}
    \caption{(a) Dispersion relations  of the superconducting leads for $B = 0.01$ T, $0.5$ T, $1.0$ T, and $2.5$~T. The corresponding extrema are indicated by red, magenta, green, and blue dots. (b) Dispersion relations of the two superconducting leads at $B = 2.5~\text{T}$ with colored lines indicating possible transitions between the gap edges. The colors denote the average value of spin-$z$ operator. (c) Numerical (dots) and analytical (lines) evaluation of these transitions as a function of $B$; the symbol $|$ marks the energies of the transitions corresponding to those in (b) while the dashed lines show halved values of the energy differences. The results are obtained for $\alpha = 50$ meVnm.}
    \label{fig:with_soc}
\end{figure*}

Before analyzing the MAR subgap structure in the presence of SOI, let us first analyze the underlying dispersion relations of the superconducting leads. Figure \ref{fig:with_soc}(a) shows four dispersion relations of the superconducting leads obtained for an increasing magnetic field $B=0.01$~T, 0.5~T, 1~T, and 2.5~T. Comparing the dispersion relation to the one obtained for the sole Zeeman interaction for $B=0.01$~T shown in Fig.~\ref{fig:without_soc}(a), we observe splitting of the bands in momentum space. When the magnetic field is increased---in contrast to the sole Zeeman interaction case---the gap in proximity to zero energy remains unaffected as a result of avoided crossings due to the spin mixing terms provided by SOI. Simultaneously, a gap in the higher (lower) energy portions of the dispersion relation opens. Finally, at $B = 2.5$~T, we observe that both the gap close to zero energy and the outermost ones are fully developed.

As the leads are semi-infinite, we can treat $k_x$ as a good quantum number and solve the Hamiltonian analytically to obtain the dispersion relation. To quantify the size of the gaps in the spectrum, we solve the analytical energy versus momentum relation for the band extrema, which allows us to approximate the energy values of the gap edges. 
The higher-energy extrema are given by 
\begin{equation}
E_{\mathrm{high}}^{\pm} = 
\sqrt{\Delta^{2} + E_{z}^{2} 
+ \frac{2\alpha^{2} m^{\ast} \mu}{\hbar^{2}}
\;\pm\; 2 \Delta E_{z}}, 
\label{eq:higher_energy}
\end{equation}
with the $+$ and $-$ branches denoted in blue and green dots in Fig. \ref{fig:with_soc}(a), respectively. 
The lower-energy extrema are described by
\begin{equation}
E_{\mathrm{low}}^{\pm} =
\left[
\Delta^{2}
+ 2\left(
  \xi_{\pm}^{2}
  - \sqrt{E_{z}^{2}\Delta^{2}+\xi_{\pm}^{4}}
\right)
\right]^{1/2}, \label{eq:lower_energy}
\end{equation}
where,  
\begin{equation}
\xi_{\pm} = \frac{\alpha^{2} m^{\ast}}{\hbar^{2}}
\,\pm\, \sqrt{E_{z}^{2}
+ \frac{2\alpha^{2} m^{\ast} \mu}{\hbar^{2}}
+ \frac{\alpha^{4} m^{\ast 2}}{\hbar^{4}}}.
\end{equation}
The extrema are marked by magenta and red dots for the $+$ and $-$ branches in Fig. \ref{fig:with_soc}(a), respectively.

Following the previously developed understanding of the position of the subgap features in conductance traces, we mark the possible transitions between the gap edges of the two superconducting leads whose dispersion relations are shown in Fig. \ref{fig:with_soc}(b). The transitions are colored according to the unique energy difference values between the gap edges. There are two unique energies for the transitions corresponding to the energy difference between the gap edge $-E^+_\mathrm{high}$ and $E^+_\mathrm{low}$ or $E^-_\mathrm{low}$, denoted with light yellow and light blue lines, respectively. The same transition energies are obtained for three transitions from the gap edges at $-E^+_\mathrm{low}$ or $-E^-_\mathrm{low}$ to $E^+_\mathrm{high}$.
There are also three unique transitions between the edges of the gap opened solely due to SOI near zero energy: two degenerate, marked with orange ($-E^-_\mathrm{low} \rightarrow E^+_\mathrm{low}$ and vice versa), and single: marked with red ($-E^-_\mathrm{low} \rightarrow E^-_\mathrm{low}$) and green ($-E^+_\mathrm{low} \rightarrow E^+_\mathrm{low}$). Finally, there is a possible transition between the outermost gap edges that we denote by the dark blue line ($-E^+_\mathrm{high} \rightarrow E^+_\mathrm{high}$).

The evolution of the energy differences between the gap edges is shown in Fig.~\ref{fig:with_soc}(c). The solid lines correspond to the differences obtained from the analytical formulas Eq.~\ref{eq:higher_energy} and Eq.~\ref{eq:lower_energy} while the symbols denote the exact gap differences extracted numerically from the energy versus momentum dependence. We observe that as the magnetic field increases, the outermost gaps are shifted apart [see Fig. \ref{fig:with_soc}(a)] and the corresponding energy differences between the gap edges increase (dark blue). On the contrary, as the magnetic field increases, it over-dominates the SOI effects, and as a result the near-zero energy gaps decrease in value [see the position of violet and red dots in Fig. \ref{fig:with_soc}(a)]. The corresponding energy differences are colored by red, orange, and green decline, as shown in Fig.~\ref{fig:with_soc}(c). It is important to note that for small magnetic fields, the gap opening around zero energy is quite stable and keeps the value of twice the superconducting gap to considerable magnetic fields. Obviously, the results for the energy differences that are different from $2\Delta$ at small fields are only tangential at zero magnetic fields, as those gaps open only in non-zero field in the presence of SOI (at zero field $E^-_\mathrm{high} = E^+_\mathrm{high}$ and $E^-_\mathrm{low} = E^+_\mathrm{low} = \Delta$).

\begin{figure}[ht!]
    \centering
    \includegraphics[width=0.6\linewidth]{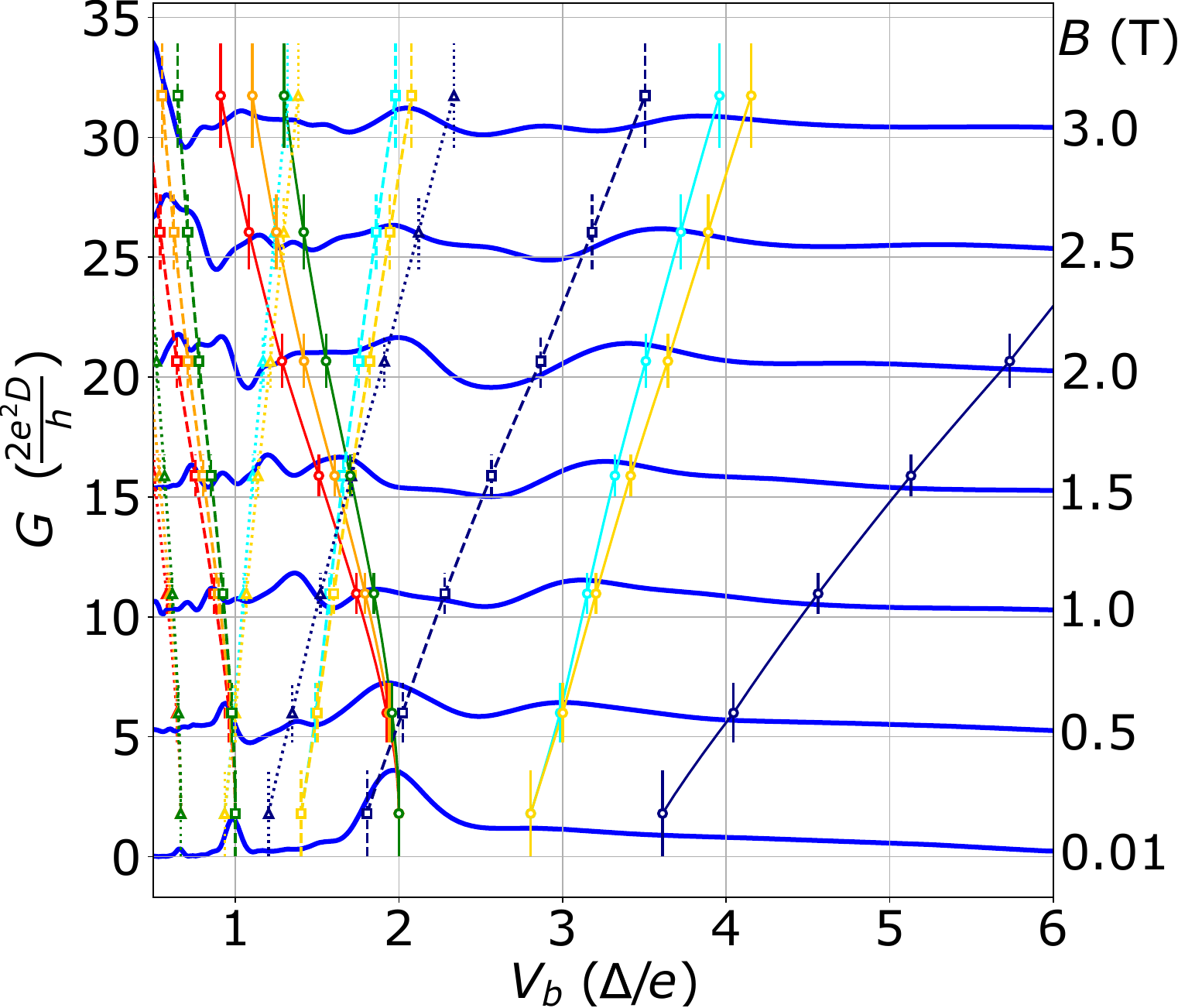}
    \caption{Conductance (blue) versus the voltage bias across the junction. The conductance traces are shifted apart by $5\cdot2e^2D/h$ for increasing magnetic field. The color lines denote the energy differences between the gap edges (solid lines) along with their ($1/n$) fractions (dashed lines).}
    \label{fig:MAR_traces}
\end{figure}

As now the energy spectrum in the superconducting leads is much richer, we expect extensive modification of MAR subgap structure in the system's  conductance traces. Figure \ref{fig:MAR_traces} presents MAR conductance versus the voltage bias $V_b$ for seven values of the magnetic field obtained for $\alpha = 50$~meVnm. For visibility purposes, the conductance curves shown in blue are offset by a value of $5\cdot2e^2D/h$. In the plot, we overlay the conductance curves by the obtained previously energy distances between the gap edges and their $1/n$ fractions. 

For nearly zero magnetic field $B = 0.01$~T, we observe that SOI does not modify the conductance trace and we obtain a usual subharmonic gap structure with jumps of conductance when the quasiparticle energy matches $2\Delta/n$ with $n$ integer. When the magnetic field increases, the energy differences between the band edges around zero energy start to decrease (see the red, orange, and green curves). Correspondingly, the peak originating at voltages corresponding to those gap edge distances moves to smaller voltages according to the evolution of the red, orange, and green curves. This is accompanied by a corresponding shift of subgap features appearing at fractions of that energy difference value, which we denote by dashed lines of the same color. Ultimately, as the gap energy differences split in a large magnetic field, we observe a corresponding splitting of the feature in the MAR conductance trace for $B>2$~T. Furthermore, as the magnetic field opens the gap that results in the energy differences marked with light yellow and light blue colors, we observe a peak developing at the corresponding voltages for magnetic fields $B \ge 0.5$ T. Additionally, we observe the development of peaks at integer fractions of those gap values denoted by dashed light blue and yellow lines.

It must be noted that since in the MAR spectrum we have transitions both at the main energy difference between the gap edges and fractions of it, and that some of the energies increase while some of them decrease as the magnitude of the magnetic field becomes larger, we might come across the situation where the first order transitions happen at the same voltage as second order ones---see the conductance trace in Fig.~\ref{fig:MAR_traces} for $B = 1.5$~T.

Finally, in the $G(V_b)$ trace, we do not observe transitions corresponding to the dark blue line. This can be understood by inspecting the spin polarization of the bands shown in Fig. \ref{fig:with_soc}(b) calculated as $\langle\sigma_z\tau_0\rangle$. The gap opened in the lowest part of the dispersion relation comes from the band, which is fully positively polarized. However, the situation is the opposite for the topmost gap, which at its top edge is constructed from a fully negatively polarized band. This spin orthogonality, which also appears at smaller magnetic fields, prohibits quasiparticle transport between the bands at the transition marked with the dark blue line. Therefore, we do not observe any conductance features at voltages that correspond to the energy difference between those gap edges. On the other hand, all other transitions are made between the bands of mixed spin polarization (which in principle reach the value of 0 at the very band edge) which allows the quasiparticle emitted from such band to be absorbed by the corresponding band in the other superconducting lead. Further analysis of the relation between the transitions and spin polarization is presented in Appendix A.

\begin{figure}[htp!]
    \centering
    \includegraphics[width=1.0\linewidth]{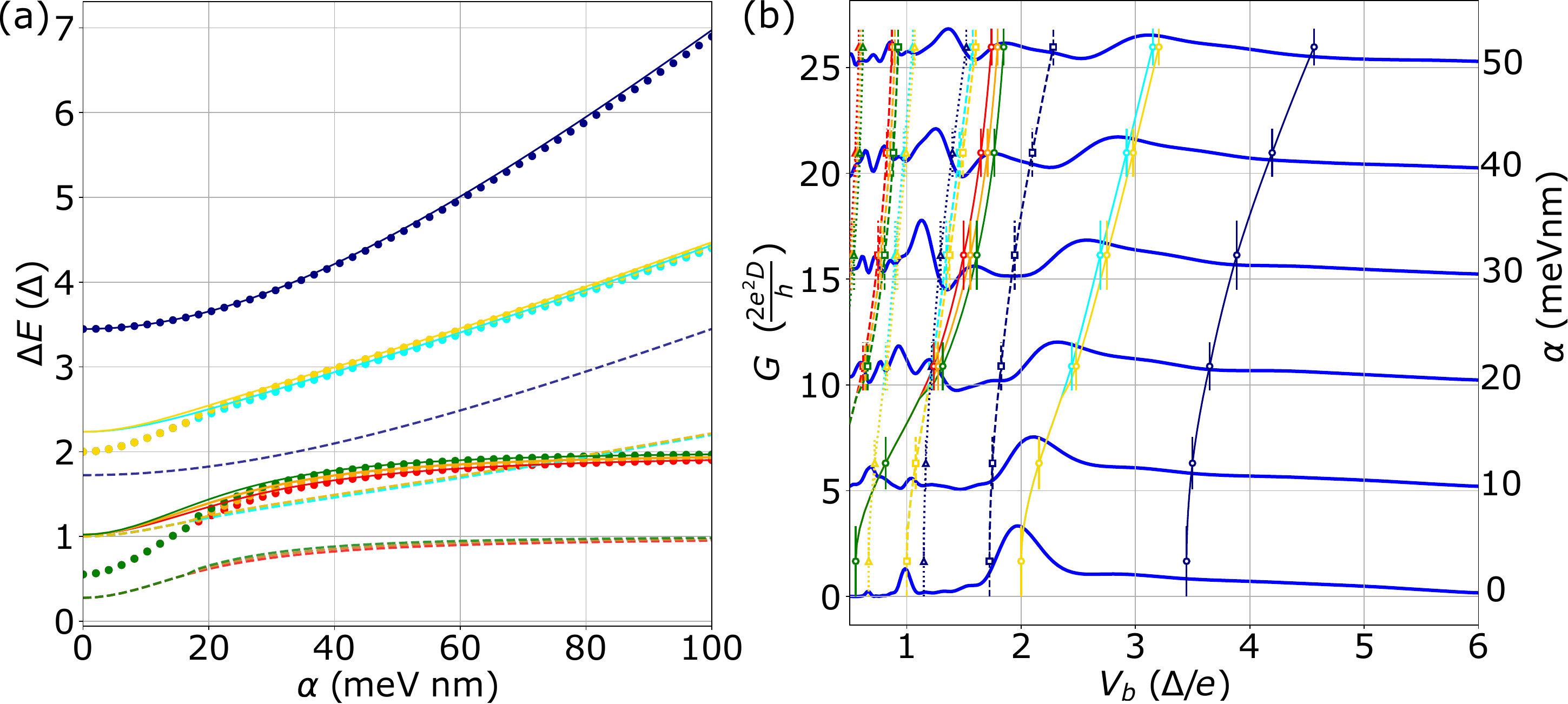} 
    \caption{(a) Numerical (dots) and analytical (lines) evaluation of possible transitions as a function of $\alpha$ at $B = 1.0$ T. (b) Conductance for $B = 1.0$ T with increasing SOI strength overlay with the numerically extracted energy distances between the gap edges (solid lines) and their $1/n$ fractions (dashed lines). The conductance traces are offset by $5\cdot2e^2D/h$ for clarity.}
    \label{fig:varied_alpha}
\end{figure}

Finally, let us now analyze how the strength of SOI modifies the MAR conductance spectra. Figure~\ref{fig:varied_alpha}(a) shows the calculated energy differences between the gap edges for varied SOI strength and a constant magnetic field of value $B = 1$~T. The dots represent the numerical solution, while the lines correspond to energy differences extracted from Eq.~\ref{eq:higher_energy} and Eq.~\ref{eq:lower_energy}. We observe a good agreement with analytically derived energy differences that, however, break at small values of $\alpha$ where Zeeman interaction effects prevail.

For negligible SOI the numerically extracted $E^-_\mathrm{low}$, $E^+_\mathrm{low}$, $E^-_\mathrm{high}$ (see Fig. \ref{fig:without_soc}(c) for $\alpha = 0$ and $B = 1$ T) coalesce into a single point. Therefore, there are only three possible energy distances between the gap edges in that situation that are equal to $2\Delta$ (yellow points), $2\Delta - 2E_z$ (green points) and finally $2\Delta + 2E_z$ (blue points) seen in Fig. \ref{fig:varied_alpha}(a) for a small $\alpha$. Only after the SOI strength increases, the full gap structure  starts to develop, fully lifting the degeneracy of $E^-_\mathrm{low}$, $E^+_\mathrm{low}$, $E^-_\mathrm{high}$ which results in the splitting of yellow and light blue lines, as well as the separation of lines colored in green, orange and red.

The corresponding evolution of the subgap features with increasing $\alpha$ is shown in Fig. \ref{fig:varied_alpha}(b) together with the numerically extracted differences between the gap edges. As SOI strength increases from zero, we observe a transition from the MAR conductance trace, which exhibits only subgap features at integer fractions of $2\Delta/e$, into one where multiple features correspond to the transition from inner to outer gap edges. Correspondingly, we observe that the main peak for $\alpha = 0$ located at $V_b = 2\Delta/e$ shifts to a higher voltage bias along with its subharmonic partner occurring close to half of that energy. The peaks at the energy difference marked with the orange, red, and green colors start to appear in the conductance trace only for $\alpha \ge 20$ meVnm. Irrespective of the value of $\alpha$, we do not find a peak corresponding to the transition with energy marked with a dark blue line as it would have to correspond to a transition between opposite-spin bands [see Fig. \ref{fig:with_soc}(b)] and here the quasiparticle spin is preserved during the transport. We, however, point out that a structure with a central peak at $V_b = 2\Delta'/e$ and two peaks at $V_b \simeq 2\Delta'/e \pm E_s$, as observed in the experiment of Ref. \cite{k8sv-lp1j} can be found in the conductance spectra for e.g. $\alpha = 40$ meVnm, but there, however, the central peak corresponds to the transitions between the gap edges at around zero energy opened by SOI (marked with red, orange, and green lines) and a transition with energy marked with light yellow and blue colors and its subharmonic partner at half of its energy. 




 




\section*{Summary and conclusions}
We studied the impact of the Zeeman interaction and spin–orbit coupling on the subharmonic gap structure in the conductance of voltage-biased Josephson junctions. Employing numerical and analytical calculations, we showed that the interplay of the Zeeman interaction, spin–orbit coupling, and superconducting pairing leads to a complex evolution of the energy gaps in the dispersion relation of the superconducting leads. Time-dependent calculations of the quasiparticle transport reveal that the emergence of these gaps strongly modifies the conductance peak structure in the conductance of the voltage-biased junction. In the absence of spin–orbit coupling, the conductance remains unaffected by the magnetic field until the superconducting gap closes at the Fermi energy. Spin mixing enabled by spin–orbit interaction opens a superconducting gap at zero energy that remains stable up to large magnetic fields and also modifies the higher-energy gaps. This, in turn, enables new transitions: the conductance is enhanced whenever quasiparticles are emitted and absorbed at the newly opened gap edges. As a result, in the conductance spectra, we observe sets of peaks that evolve toward either smaller or larger voltages as the magnetic field is increased, consistent with the energies extracted from the dispersion relations of the leads. Importantly, higher-order multiple Andreev reflection processes can cause coalescence of the peaks as a result of transport between different gap edges. Finally, within our single-channel model, we demonstrate that the visibility of the conductance peaks is governed by spin selection rules: quasiparticle transport is suppressed between bands with opposite spin polarizations and enhanced between bands with aligned spin polarizations.

\appendix
\section{Spin-dependent transitions}
\begin{figure}[htp!]
    \centering
    \includegraphics[width=0.7\linewidth]{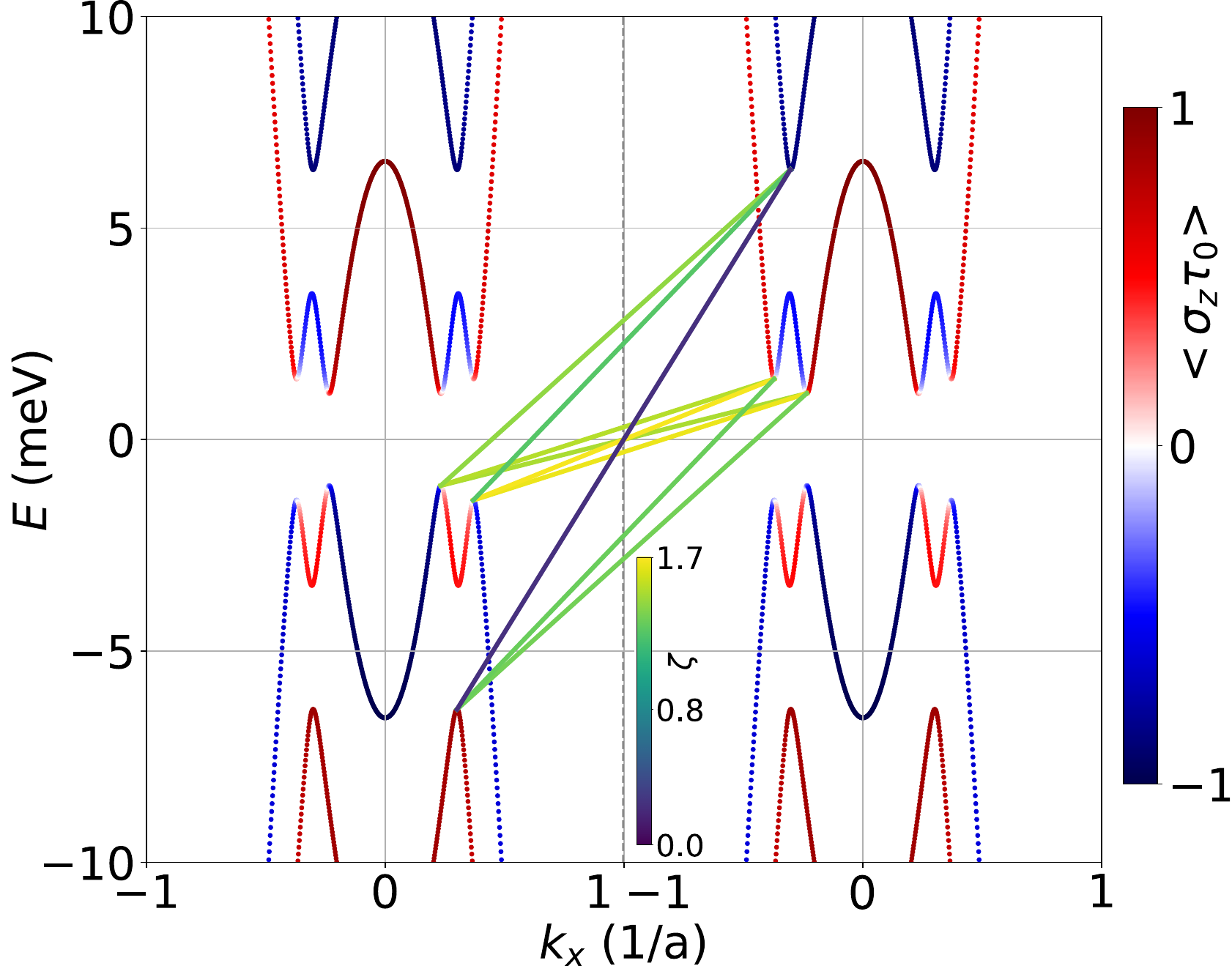} 
    \caption{Dispersion relations of the two superconducting leads at $B = 2.5$ T with colored lines of the given transitions indicating the relative spin alignment ($\zeta$) of the states.}
    \label{fig:spin_similarity}
\end{figure}

To further quantify the relative spin alignment of the states connected by a given transition, we define the factor
\begin{equation}
\zeta = 2 - \left| \langle \sigma_z \tau_0 \rangle_j - \langle \sigma_z \tau_0 \rangle_i \right| ,
\end{equation}
where $\langle \sigma_z \tau_0 \rangle_i$ and $\langle \sigma_z \tau_0 \rangle_j$ denote the spin expectation values of the eigenstates at the band maxima and minima, respectively, evaluated at the corresponding momenta connected by the transition lines. The color of these lines encodes the magnitude of this spin-alignment factor, as shown in Fig. \ref{fig:spin_similarity}.

We find that for the smaller gaps, the spin expectation values are small in magnitude, indicating strongly spin-mixed states. Consequently, the factor $\zeta$ has values in the range $\sim 1.5\text{--}1.7$, reflecting a relatively weak spin mismatch between the connected states. In contrast, for the largest gap, the involved states are strongly spin-polarized and exhibit nearly opposite spin orientations. As a result, the factor approaches zero, indicating a pronounced spin mismatch.

\section{Effect of junction transparency on conductance}

\begin{figure}[htp!]
    \centering
    \includegraphics[width=1.0\linewidth]{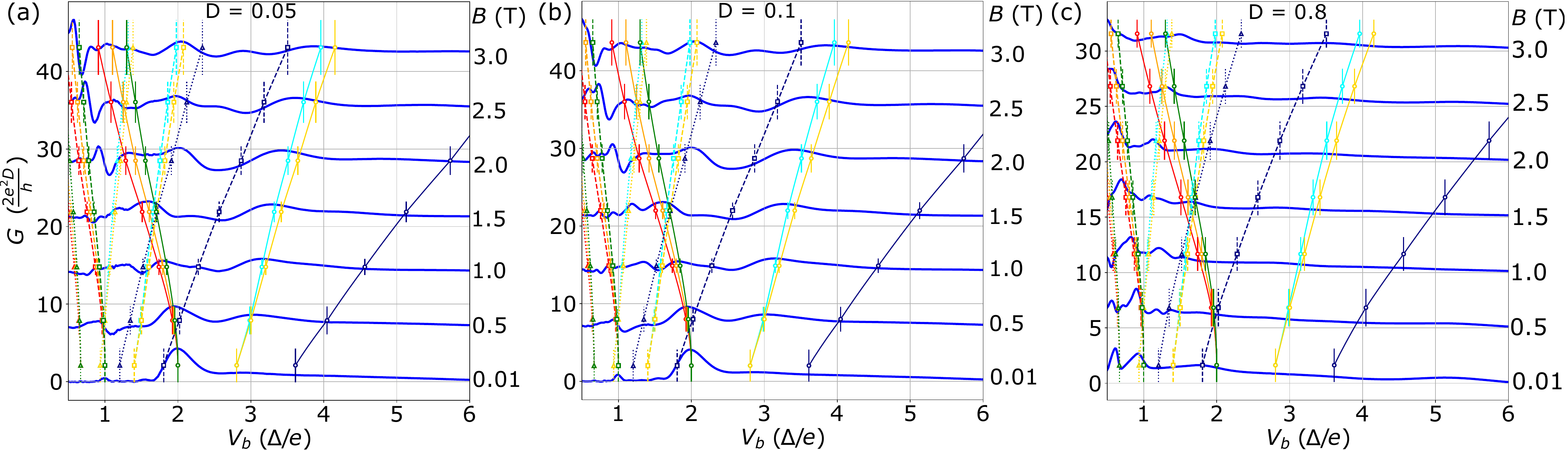} 
    \caption{Conductance (blue) as a function of bias voltage for different junction transparencies: (a) $D = 0.05$, (b) $D = 0.1$, and (c) $D = 0.8$. The colored lines indicate the energy differences between the gap edges (solid lines) and their subharmonic fractions $1/n$ (dashed lines).}
    \label{fig:different_transparency}
\end{figure}

To assess the robustness of our conclusions with respect to junction transparency, we perform additional calculations for several values of transparency, namely $D = 0.05$, $0.1$, and $0.8$, in a single-channel junction. Representative results are shown in Fig.~\ref{fig:different_transparency}.

Although the overall conductance amplitude and line shape depend on $D$, the qualitative features central to our analysis, in particular the positions and evolution of the MAR-related structures induced by the modified band structure, remain robust throughout this range. In particular, for low and intermediate transparencies ($D = 0.05$ and $D = 0.1$), the additional MAR-related subharmonic features persist and follow the same trends as a function of bias voltage and magnetic field, as shown in Fig.~\ref{fig:different_transparency}(a,b).

At high transparencies (e.g., $D \gtrsim 0.8$), the subharmonic gap structure in single-channel junctions becomes less pronounced (Fig.~\ref{fig:different_transparency}(c)). As discussed in Ref.~\cite{PhysRevLett.75.1831}, increasing transparency leads to a progressive broadening of the MAR features and a smoother evolution of the current--voltage characteristics as $D \to 1$. In this regime, the structures in the differential conductance may appear as dips rather than peaks, consistent with the behavior reported in Ref.~\cite{PhysRevApplied.7.034029}.


\bibliography{references}

\begin{thebibliography}{10}
\bibitem{B0comment}
\bibinfo{note}{For system Hamiltonian including spin degree of freedom the solutions for $B = 0$ were not stable numerically, therefore we resort to a case of small $B = 0.01$ T, which has neglible effect in terms of Zeeman interaction.}
\end{thebibliography}








\section*{Funding}
This work was supported by the National Science Center (NCN), Poland, Agreement No. UMO-2020/38/E/ST3/00418 and partially by the program 'Excellence initiative - research university' for AGH University. We gratefully acknowledge the Polish high-performance computing infrastructure PLGrid (HPC Center: ACK Cyfronet AGH) for providing computer facilities and support within the computational grant no. PLG/2025/018486.

\section*{Author contributions statement}
D.K. conducted all the calculations presented in this work. M.P.N. conceived the main idea behind this work. D.K. and M.P.N. wrote the first version of the manuscript. D.K., J.H.C., A.B. and M.P.N. participated in the analysis and interpretation of the obtained results. All authors reviewed the manuscript.

{\section*{Code Availability and Reproducibility}
The code used for the calculations presented in this work is available in an online repository~\cite{kuiri_2026_19204942} including Jupyter notebooks with the code and the corresponding precomputed data files along with an instruction how to reproduce the data.

\end{document}